\documentclass[a4paper,english,twocolumn,aps,prd,nofootinbib]{revtex4}
\usepackage[T1]{fontenc}
\usepackage[latin9]{inputenc}
\usepackage{color}
\usepackage{babel}
\usepackage{amsmath}
\usepackage{amssymb}
\usepackage{graphicx}

\usepackage{hyperref}
\hypersetup{
  colorlinks=true,        
  linkcolor=blue,         
  citecolor=cyan,         
}
\usepackage{breakurl}
\usepackage{graphicx}
\usepackage{epsf}
\usepackage{epsfig}
\usepackage{times}

\newcommand{\cf}{cf.~}
\newcommand{\ie}{i.e.,~}
\newcommand{\eg}{e.g.,~}

\makeatother

\begin{document}

\title{Bjorken flow in one-dimensional relativistic magnetohydrodynamics
with magnetization}

\author{Shi Pu$^{1}$, Victor Roy$^{1}$, Luciano Rezzolla$^{1,2}$, and Dirk H.\
Rischke$^{1}$}

\affiliation{$^{1}$ Institute for Theoretical Physics, Goethe University,\\
Max-von-Laue-Str.\ 1, 60438 Frankfurt am Main, Germany}

\affiliation{$^{2}$ Frankfurt Institute for Advanced Studies, Goethe
  University,\\ Ruth-Moufang-Str.\ 1, 60438 Frankfurt am Main, Germany}

\begin{abstract}
We study the one-dimensional, longitudinally boost-invariant motion of an
ideal fluid with infinite conductivity in the presence of a transverse
magnetic field, \ie in the ideal transverse magnetohydrodynamical limit.
In an extension of our previous work Roy et al., [Phys. Lett. B 750, 45 (2015)], we consider the
fluid to have a non-zero magnetization. First, we assume a constant
magnetic susceptibility $\chi_{m}$ and consider an ultrarelativistic
ideal gas equation of state. For a paramagnetic fluid (\ie with
$\chi_{m}>0$), the decay of the energy density slows down since the fluid
gains energy from the magnetic field. For a diamagnetic fluid (\ie with
$\chi_{m}<0$), the energy density decays faster because it feeds energy
into the magnetic field. Furthermore, when the magnetic field is taken to
be external and to decay in proper time $\tau$ with a power law
$\sim\tau^{-a}$, two distinct solutions can be found depending on the
values of $a$ and $\chi_m$. Finally, we also solve the ideal
magnetohydrodynamical equations for one-dimensional Bjorken flow with a
temperature-dependent magnetic susceptibility and a realistic equation of state given
by lattice-QCD data. We find that the temperature and energy density decay
more slowly because of the non-vanishing magnetization. For values of the
magnetic field typical for heavy-ion collisions, this effect is, however,
rather small. Only for magnetic fields which are about an order of
magnitude larger than expected for heavy-ion collisions, the system is
substantially reheated and the lifetime of the quark phase might be
extended.
\end{abstract}
\maketitle

\section{Introduction}

It has recently been pointed out that extremely strong magnetic fields
are produced in relativistic heavy-ion collisions. In general, their
magnitde grows approximately linearly with the centre-of-momentum energy
of the colliding nucleons \cite{Bzdak:2011yy,Deng:2012pc,Li:2016tel}, and reaches
$B\sim10^{18}-10^{19}$ G in Au-Au collisions at $\sqrt{s_{NN}} = 200$ GeV. In
these collisions a new form of hot and dense nuclear matter is created,
commonly known as quark-gluon plasma (QGP) \cite{Gyulassy:2004zy}. In the
QGP, quarks are deconfined and chiral symmetry is restored, such that
quarks are (approximately) massless and have definite chirality and
helicity. Thus, in the presence of very strong magnetic fields, quarks
will be polarized and will move preferentially in the direction parallel
(or antiparallel) to the magnetic field. Therefore, if the numbers of
left- and right-handed quarks are not equal, a net charge current will be
induced, a phenomenon known as ``chiral magnetic effect'' (CME)
\cite{Kharzeev:2007jp,Fukushima:2008xe}. As well as charge currents, a
chiral current can also be induced by the magnetic field and gives rise
to the ``chiral separation effect'' (CSE). Combining these two effects, a
density wave is expected to be induced by magnetic fields, called
``chiral magnetic wave'' (CMW) \cite{Kharzeev:2010gd}, which might break
the degeneracy between the elliptic flows of $\pi^{\pm}$
\cite{Burnier:2011bf}. Recently, it has been found that these phenomena
can be interpreted in the language of the Berry phase and effective
chiral kinetic equations, which can be obtained by the
path-integral method \cite{Stephanov:2012ki,Chen:2013iga,Chen:2014cla},
Hamiltonian approaches \cite{Son:2012wh,Son:2012zy}, and quantum kinetic
theory via Wigner functions \cite{Gao:2012ix,Chen:2012ca}; for reviews
and additional references see Refs.\ \cite{Bzdak:2012ia,
  Kharzeev:2013ffa, Kharzeev:2015kna, Pu:Review2015}.

A possible origin of CME, CSE, and CMW is the chiral anomaly, which is
topologically invariant. Thus, these effects are not expected to be
significantly modified when considering interactions between particles.
In contrast, there are also several interesting phenomena dominated by
the interactions. For example, instead of a strong magnetic field, a
chiral current and density wave can also be induced by an electric field,
the so-called ``chiral electric separation effect'' (CESE)
\cite{Huang:2013iia,Pu:2014cwa,Jiang:2014ura,Pu:2014fva}. Similarly,
adding an electric field perpendicular to the magnetic field, a chiral
Hall current is expected, called ``chiral Hall-separation effect'' (CHSE)
\cite{Pu:2014fva}, which might cause an asymmetric charge and chirality
distribution in rapidity. These phenomena have drawn considerable
attention within the study of hot and dense matter under the influence of
strong magnetic fields.

A very popular and successful tool to describe heavy-ion collision
dynamics is relativistic hydrodynamics [see, \eg
  Refs.\ \cite{Romatschke2007, Luzum2008, Song2008b, Song:2008si,
    Schenke:2011bn, Roy:2012jb, Niemi:2012ry, Rezzolla2013}]. In order to
study some of the phenomena mentioned above within a hydrodynamical
approach, the latter needs to be extended to include magnetic fields, \ie
one has to develop and apply a relativistic magnetohydrodynamical (MHD)
framework. Although the magnetic field created in heavy-ion collisions
rapidly decays in the vacuum \cite{Kharzeev:2007jp} and might become very
small even before the system reaches local thermal equilibrium, some
recent studies have shown that its decay might be substantially delayed
in the presence of an electrically conducting medium
\cite{Gursoy:2014aka, Zakharov:2014dia, Tuchin:2013apa}. Thus, it is
still presently unclear whether the effect of magnetic fields on the
dynamics of a heavy-ion collision can be neglected or needs to be taken
properly into account.

In analogy with our previous work \cite{Roy:2015kma} (hereafter paper I),
we define the dimensionless quantity, $\sigma=B^{2}/e$, to measure the
relative importance of the magnetic field, with $B$ being the magnitude
of the magnetic field (measured in units of GeV$^2$) and $e$ the energy
density of the fluid. Clearly, for regions where $\sigma\gtrsim1$, the
effect of the magnetic fields cannot be ignored. Interestingly, in a
typical mid-central Au-Au collision (\eg with impact parameter $\sim10$
fm at $\sqrt{s_{NN}}=200$ GeV), the average magnetic field is $\sim10
\,m_{\pi}^{2}$ \cite{Bzdak:2011yy,Deng:2012pc}, with $m_{\pi}$ the pion
mass, and the energy density is $\sim10$ GeV fm$^{-3}$, thus giving
$\sigma\sim 0.5$. Furthermore, event-by-event simulations show that in certain events
$\sigma$ could be even much larger than $1$ in certain regions
\cite{Roy:2015coa}. Therefore, it is still very important to investigate
relativistic MHD, ideally with a numerical code solving the equations
in 3+1 dimensions.

Before starting to investigate MHD numerically, it is worth while to
search for analytic solutions in some simple, but nevertheless realistic
test cases. In paper I we have considered one-dimensional, longitudinally
boost-invariant Bjorken flow \cite{Bjorken:1982qr} with a transverse
magnetic field and in ideal MHD, \ie for infinite electrical conductivity
and without dissipative effects. Quite remarkably, we found that under
these conditions the decay of the energy density is the same as in the
case without magnetic field. The reason is that, for a transverse
magnetic field, $B/s$ (where $s$ is the entropy density) is conserved and
the magnetic field is advected with the fluid, as a manifestation of the 
"frozen-flux theorem" well-known from astrophysics and plasma
physics \cite{Rezzolla2013, Landau}.

In paper I we have neglected the effect of a nonzero magnetization of the
QGP. In this work, we extend our previous study to include a nonzero
magnetization for the scenario of a pure Bjorken flow. The final goals
are those of improving the theoretical understanding of the MHD evolution
of the QGP, but also of determining how the magnetization of the QGP may
have an influence on measurable quantities, as recently suggested in
Ref.\ \cite{Bali:2013owa} in relation to the value of the elliptic flow
parameter $v_{2}$.

Within a linear approximation, the magnetization effect can be described
through the magnetic susceptibility $\chi_{m}$, which is the ratio of the
magnetic polarization to the magnetic field\footnote{Note that usually
  the magnetic susceptibility is related to the quark condensate $\langle
  \overline{\psi}\sigma_{\mu\nu}\psi \rangle$ \cite{Ioffe:1983ju}, and
  has been studied in a number of works, see \eg
  Refs.\ \cite{Bali:2012jv,Buividovich:2009ih} for lattice QCD,
  Ref.\ \cite{Frasca:2011zn} for other theoretical studies. Here, we
  concentrate on the \emph{full} magnetic susceptibility, which is
  defined through a derivative of the pressure (grand canonical
  potential); more details will be discussed in Sec.\ \ref{sec:Ideal
    MHD}.}. Interestingly, numerous studies, \eg from lattice QCD
\cite{Bali:2013owa, Bonati:2013lca, Bonati:2013qra, Levkova:2013qda,
  Bonati:2013vba, Orlovsky:2014cta, Bali:2014kia} and from perturbative
QCD \cite{Kamikado:2014bua}, using the Sakai-Sugimoto model
\cite{Bergman:2008sg}, the functional renormalization group
\cite{Kamikado:2014bua}, or other models \cite{Cherman:2008eh,
  Steinert:2013fza, Kabat:2002er, Orlovsky:2014cta, Anber:2013tra,
  Tawfik:2014hwa}, have all suggested that in a confined phase (hadron
phase) the medium is diamagnetic, \ie with $\chi_{m}<0$, while in the
deconfined QGP phase it is paramagnetic, \ie with $\chi_{m}>0$. As a
result, by simply choosing different signs of $\chi_{m}$, we can
investigate the dynamics of different phases with nonzero magnetization.

In principle, the magnetic susceptibility $\chi_{m}$ of the QGP is a
function of the temperature and of the magnitude of the magnetic
field. In practice, however, the variation of $\chi_{m}$ is very small in
the experimentally accessible region of temperatures and magnetic
fields. For example, from lattice-QCD calculations
\cite{Bali:2013owa,Bonati:2013qra} [see also
  Ref.\ \cite{Orlovsky:2014cta}], for $(eB)^{2}=0.007-0.2$ GeV$^{2}$ and
$T=100-350$ MeV, the susceptibility is in the range
$0\lesssim\chi_{m}\lesssim0.05$ [note that a prefactor $4\pi \alpha$ is
  required to convert the values of Refs.\ \cite{Bali:2013owa,
    Bonati:2013qra, Orlovsky:2014cta} from SI to natural units]. In view
of this and for the sake of simplicity, we will first assume a constant
$\chi_{m}$ and then investigate the modifications of the dynamics when
$\chi_m$ is assumed to depend on the temperature.


This paper is organized as follows. In Sec.\ \ref{sec:Ideal MHD}, we
introduce the ideal-MHD framework with nonzero magnetization. We also
discuss the conservation equations and apply them to 
Bjorken flow in the presence of a magnetic field. In
Sec.\ \ref{sec:Energy_ideal}, we obtain the temporal evolution of the
energy density in ideal transverse MHD with magnetization. As a useful
comparison, we consider in Sec.\ \ref{sec:power law} the decay of the
energy density in the presence of an external magnetic field undergoing a
a power-law decay in proper time. In
Sec.\ \ref{sec:Temperature-dependent-magnetic} we solve the MHD equations
numerically for a temperature-dependent magnetic susceptibility and a
realistic equation of state (EOS). Finally, we summarize and conclude in
Sec.\ \ref{sec:Conclusion}.

Throughout this work, we work in a flat spacetime with the metric tensor
$g_{\mu\nu}=\eta_{\mu\nu}={\rm diag}\{+,-,-,-\}$, so that the fluid
four-velocity $u^\mu$ satisfies $u^{\mu}u_{\mu}=1$ and the orthogonal
projector to the fluid four-velocity is defined as
$\Delta^{\mu\nu}=g^{\mu\nu}-u^{\mu}u^{\nu}$. We will also use the
Levi-Civita tensor with $\epsilon^{0123}=-\epsilon_{0123}=1$. Note that
in this convention, contracting the indices of two Levi-Civita tensors
will have an additional minus sign, \eg $\epsilon^{\mu\nu\alpha\beta}
\epsilon_{\mu\nu\rho\sigma} =-(g_{\rho}^{\alpha}g_{\sigma}^{\beta} -
g_{\sigma}^{\alpha}g_{\rho}^{\beta})$.

\section{Ideal MHD with magnetization}
\label{sec:Ideal MHD}

\subsection{Covariant form of the MHD equations}

In this section, we give a brief introduction to the Lorentz-covariant
form of ideal MHD with nonzero magnetization. More details can be found
in Refs.\ \cite{Caldarelli:2008ze, Gedalin:PRE1995, Huang:2009ue,
  Groot:Maxwell}, which use the same convention as this work, or in
Refs.\ \cite{Giacomazzo:2005jy,Giacomazzo:2007ti}, which use instead a
$+2$ signature for the metric. 

We start by recalling that in the presence of an electromagnetic field,
the (total) energy-momentum tensor of an ideal fluid can be decomposed
into two parts
\begin{equation}
\label{eq:total_EMT_01}
T^{\mu\nu} = T_{_{M}}^{\mu\nu} + T_{_{EM}}^{\mu\nu}\,,
\end{equation}
where 
\begin{equation}
T_{_{EM}}^{\mu\nu} = -F^{\mu\lambda}F_{\lambda}^{\nu} +
\frac{1}{4}g^{\mu\nu}F^{\alpha\beta}F_{\alpha\beta}\,,
\end{equation}
is the contribution from the electromagnetic fields and
\begin{equation} 
\label{T_M}
T_{_{M}}^{\mu\nu} =
eu^{\mu}u^{\nu}-p\Delta^{\mu\nu}-\frac{1}{2}(M^{\mu\lambda}F_{\lambda}^{\;\nu}
+ M^{\nu\lambda}F_{\lambda}^{\;\mu})\,,
\end{equation}
refers to the matter part, with $p$ being the thermodynamic pressure
\cite{Rezzolla2013}. The polarization tensor $M^{\mu\nu}$ can be
expressed directly through the derivatives of the grand canonical
potential $\Omega$ with respect to the electromagnetic field (or Faraday tensor)
$F^{\mu\nu}$ as 
\begin{equation}
M^{\mu\nu} \equiv -\frac{\partial\Omega(T,\mu,B)}
{\partial F_{\mu\nu}}\,,
\end{equation}
where $T,\mu,B$ are the temperature, the chemical potential, and the
strength of the magnetic field, respectively. In the weak-field limit,
the terms $\sim M^{\mu\nu}$ in Eq.\ (\ref{T_M}) can be neglected.

The equations for the conservation of the (total) energy and momentum
are given simply by
\begin{equation}
\label{eq:EM_con_01}
\partial_{\mu}T^{\mu\nu} = 0\,, 
\end{equation}
and reduce, with the help of Maxwell's equations, to the well-known form
\begin{equation}
\partial_{\mu}T_{_{M}}^{\mu\nu} =
-\partial_{\mu}T_{_{EM}}^{\mu\nu} = F^{\mu\lambda}j_{\lambda}\,, 
\end{equation}
with $j^{\mu}$ being the charge current. Without loss of generality, the
electromagnetic field tensor can be decomposed as
\begin{equation}
\label{eq:F_01}
F^{\mu\nu} = E^{\mu}u^{\nu}-E^{\nu}u^{\mu} +
\epsilon^{\mu\nu\alpha\beta}u_{\alpha}B_{\beta}\,,
\end{equation}
with  
\begin{equation}
E^{\mu} \equiv F^{\mu\nu}u_{\nu}\,,\qquad B^{\mu} \equiv
\frac{1}{2}\epsilon^{\mu\nu\alpha\beta}u_{\nu}F_{\alpha\beta}\,.
\end{equation}
In the local rest frame of the fluid, where $u^{\mu} =
(1,\boldsymbol{0})$, the spatial components of $E^\mu$ represent the
electric-field three-vector, while those of $B^\mu$ refer to
magnetic-field three-vector. We can next introduce the in-medium
field-strength tensor $H^{\mu\nu}\equiv F^{\mu\nu}-M^{\mu\nu}$ and
similarly decompose $H^{\mu\nu}$ and $M^{\mu\nu}$ as
\begin{eqnarray}
M^{\mu\nu} & = & P^{\nu}u^{\mu}-P^{\mu}u^{\nu} +
\epsilon^{\mu\nu\alpha\beta}u_{\alpha}M_{\beta},\nonumber \\ H^{\mu\nu} &
= & D^{\mu}u^{\nu}-D^{\nu}u^{\mu} +
\epsilon^{\mu\nu\alpha\beta}u_{\alpha}H_{\beta}\,,
\end{eqnarray}
where 
\begin{align}
P^{\mu} & \equiv   -M^{\mu\nu}u_{\nu}\,,  &\qquad 
M^{\mu} & \equiv 
\frac{1}{2}\epsilon^{\mu\nu\alpha\beta}u_{\nu}M_{\alpha\beta}\,,
\\ D^{\mu} & \equiv   H^{\mu\nu}u_{\nu}\,, & \qquad 
H^{\mu} & \equiv 
\frac{1}{2}\epsilon^{\mu\nu\alpha\beta}u_{\nu}H_{\alpha\beta}\,.
\end{align}
In the local rest frame the spatial components $\boldsymbol{P},
\boldsymbol{M}$ are the electric and magnetic polarization vectors,
respectively. Similarly, the spatial components $\boldsymbol{D}$ and
$\boldsymbol{H}$ are the electric displacement field and magnetic field
intensity, respectively. Note that the minus sign in the definition of
$P^{\mu}$ is coming from the fact that the polarization field in the
medium points in the direction opposite to the external electric
field. Hereafter, we will use $M^{\mu\nu}$ instead of $H^{\mu\nu}$ to
discuss the magnetization effects.

Besides the direction given by the fluid velocity, the magnetic field
singles out another special direction in the system, which we associate
with the spacelike unit vector\footnote{For completeness, and to avoid
  confusion, we note that in general relativistic MHD a different
  notation is normally adopted [see, \eg Refs.\
  \cite{Giacomazzo:2005jy,Giacomazzo:2007ti}]. First, the signature is
  spacelike, \ie $g_{\mu\nu} = \mbox{\ensuremath{\left(-, + , + , +
      \right)}}$. Second, $b^{\mu}$ is defined as the magnetic field
  four-vector measured in a comoving frame, while $\boldsymbol{B}$ still
  represents the magnetic field three-vector measured by an Eulerian
  observer. Third, when $M=0$, the energy-momentum tensor takes the form
  $T^{\mu\nu} = (e + p + b^{2})u^{\mu}u^{\nu} + \left(p +
  b^{2}/2\right)g^{\mu\nu}-b^{\mu}b^{\nu}$ [\cf Eq. \eqref{eq:EMT_01}].}
\begin{equation}
\label{eq:def_b}
b^{\mu} \equiv \frac{B^{\mu}}{B}\,,
\end{equation}
where
\begin{equation}
B \equiv \sqrt{-B^{\mu}B_{\mu}}\,, \qquad\qquad
b^{\mu}b_{\mu} = -1\,.
\end{equation}
In a linear approximation we can rewrite $M^{\mu}$ as
\begin{equation}
\label{eq:susceptibility_01}
M^{\mu}\equiv Mb^{\mu} = \chi_{m}B^{\mu}\,, 
\end{equation}
where
\begin{equation}
M \equiv \chi_{m}B =
\sqrt{-M^{\mu}M_{\mu}}\,.
\end{equation}
Since $B^{\mu}B_{\mu}$ and $M^{\mu}M_{\mu}$ are Lorentz scalars, $B$ and
$M$ represent the magnitude of the magnetic field and the magnetization
in the local rest frame of the fluid, respectively. In addition,
$\chi_{m}$ is the magnetic susceptibility, which in principle is a
function of temperature and of the magnetic field strength\footnote{From
  the definition (\ref{eq:susceptibility_01}) and
  Eq.\ (\ref{eq:M_canonical_01}), $\chi_{m}$ can be obtained through the
  expansion of the pressure or of the grand canonical potential in the
  weak-field case, $p(T,B) = p(T,0) + \chi_{m}B^{2}/2 +
  \mathcal{O}(B^{4})$. Unfortunately, $\chi_{m}$ is divergent and is
  related to the electric charge renormalization in QED
  \cite{Bali:2013owa, Kamikado:2014bua, Bali:2014kia}. As a result, it is
  common to define a renormalized magnetic susceptibility,
  $\tilde{\chi}_{m}(T)\equiv\chi(T)-\chi(0)$ \cite{Kamikado:2014bua}. In
  this work, we only consider $\chi_m$ as a free parameter, avoiding the
  difficulties of possible singularities.}. As mentioned in the
Introduction, we will first consider the case in which $\chi_{m}$ is a
constant and subsequently the case in which $\chi_{m}$ has a dependence
on temperature.

Although event-by-event simulations of heavy-ion collisions show that the
electric field in the laboratory frame could be as large as the magnetic
field, we restrict our attention here to the ideal transverse MHD limit.
In such a limit, the electric conductivity is assumed to be infinite
$\kappa\rightarrow\infty$ (\ie the medium is a perfectly conducting
plasma), thus requiring the electric field to vanish in the comoving
frame although the charge current $j^{\mu} = \kappa E^{\mu}$ can be
finite. Alternatively, this can be seen as a condition on the electric
and magnetic fields in the lab frame, $\boldsymbol{E}, \boldsymbol{B}$,
which are related via the simple algebraic relation $\boldsymbol{E} +
\boldsymbol{v}\times\boldsymbol{B} = 0$, with $\boldsymbol{v}$ being the
three-velocity of the fluid, also in the lab frame. As a result, in the
ideal-MHD limit and a linear approximation, the electric polarization
vector $P^{\mu}$ and the electric displacement field $D^{\mu}$ can be
neglected in Eq.\ (\ref{eq:total_EMT_01}).

Under these assumptions, Maxwell's equations simplify considerably, and
the first couple of Maxwell equations is given by
\begin{equation}
\epsilon^{\nu\mu\alpha\beta}\partial_{\nu}F_{\alpha\beta} = 0\,,
\end{equation}
or, equivalently
\begin{equation}
\label{eq:Maxwell_02}
\partial_{\nu}(B^{\mu}u^{\nu}-B^{\nu}u^{\mu}) = 0\,.
\end{equation}
Alternative expressions can be obtained after contracting
Eq.\ (\ref{eq:Maxwell_02}) with $B_{\mu}$ to obtain
\begin{equation}
\frac{1}{2}(u^{\alpha}\partial_{\alpha})B^{2} + B^{2}\partial_{\alpha} u^{\alpha} +
B^{2}b^{\mu}b^{\nu}\partial_{\nu}u_{\mu} = 0\,,\label{eq:Maxwell_03}
\end{equation}
or when considering Eq.\ (\ref{eq:Maxwell_02}) in the local rest frame,
in which case it reduces to the following well-known Maxwell equations
\begin{eqnarray}
\label{eq:pre_Max_01}
\nabla\cdot\boldsymbol{B} & = & 0\,,\nonumber \\ 
\partial_{t}\boldsymbol{B} & = &
-\nabla\times\boldsymbol{E} =
\nabla\times(\boldsymbol{v}\times\boldsymbol{B})\,.
\end{eqnarray}
Similarly, the second couple of Maxwell equations is instead given by
\begin{equation}
\label{eq:Maxwell_04}
\partial_{\mu}H^{\mu\nu} = j^{\nu}\,,
\end{equation}
which can be seen as constraint conditions for the charge density and
external fields. Because these equations will not be part of our
following discussion, we discuss them only briefly. In non-relativistic
plasma physics, the charge density $j^{0}$ and its time derivative
$\partial_{t}j^{0}$ are negligible when compared to the other terms in
Eqs.\ (\ref{eq:Maxwell_04}), so that the latter reduce to Ampere's Law,
\ie $\nabla\times\boldsymbol{B} = \boldsymbol{j}$ and to
$\nabla\cdot\boldsymbol{j} = 0$. In relativistic ideal MHD, on the other
hand, the fluid is perfectly conducting and $E^{\mu},D^{\mu},P^{\mu}$
vanish in the co-moving frame of the fluid, so that
Eq.\ (\ref{eq:Maxwell_04}) becomes
\begin{equation}
\label{eq:Maxwell_05}
\epsilon^{\mu\nu\alpha\beta}\partial_{\mu}[u_{\beta}(B_{\alpha}-M_{\alpha})]
= j^{\nu}\,.
\end{equation}
%

Let us now turn to the thermodynamical relations that will be useful in
the remainder of this work. We recall that for a perfect fluid in
thermodynamical equilibrium [see, \eg Refs.\ \cite{Caldarelli:2008ze,
  Huang:2009ue, Bali:2014kia, Landau:EM, Rezzolla2013}]
\begin{equation}
\label{eq:thermo_relation_01}
e + p = Ts + \mu n\,,
\end{equation}
where $n$ is the baryon number density and $\mu$ the associated chemical
potential. Furthermore, from the definition of the grand canonical
potential $\Omega(T,\mu,B) \equiv -pV$, the magnitude of the magnetic
polarization vector is given by
\begin{equation}
M = -\frac{1}{V}\left.\frac{\partial\Omega}{\partial B}\right|_{T,\mu} =
\left.\frac{\partial p}{\partial B}\right|_{T,\mu}\,.
\label{eq:M_canonical_01}
\end{equation}
This implies that 
\begin{equation}
\label{eq:Gibbs_dp_02}
dp = sdT + nd\mu + MdB\,,
\end{equation}
and thus that
\begin{equation}
\label{eq:Gibbs_de_02}
de = Tds + \mu dn-MdB\,.
\end{equation}
%

In ultrarelativistic heavy-ion collisions, the net baryon number density
and the chemical potential are vanishingly small at mid-rapidity,
therefore, for the sake of simplicity, we will only consider the case of
zero baryon chemical potential throughout this work. Then, the
thermodynamic relations \eqref{eq:thermo_relation_01},
\eqref{eq:Gibbs_dp_02}, and \eqref{eq:Gibbs_de_02} reduce to
\begin{eqnarray}
\label{eq:thermo_relation_01n}
e + p &  =  & Ts\,, \\
\label{eq:Gibbs_dp_02n}
dp &  =  & sdT + MdB\,, \\
\label{eq:Gibbs_03}
de &  =  & Tds-MdB\,,
\end{eqnarray}
while the sound speed is defined as
\begin{equation}
\label{eq:cs2}
c^2_s \equiv \left.\frac{\partial p}{\partial e}\right|_{s,B}\,.
\end{equation}



\subsection{Conservation equations and EOS}

Inserting the definition of the Faraday tensor\ (\ref{eq:F_01}) into the
expression of the (total) energy-momentum tensor
Eq.\ (\ref{eq:total_EMT_01}), in ideal MHD the latter can expressed as
\cite{Gedalin:PRE1995,Huang:2009ue} 
\begin{eqnarray}
\label{eq:EMT_01}
T^{\mu\nu} &=& (e + p-MB + B^{2})u^{\mu}u^{\nu}-
(p-MB + \frac{1}{2}B^{2})g^{\mu\nu} \nonumber\\
&\!\phantom{=}\!& + (MB-B^{2})b^{\mu}b^{\nu}\,,
\end{eqnarray}
where $M,B$, and $b^{\mu}$ are defined in Eqs.\ (\ref{eq:def_b}),
(\ref{eq:susceptibility_01}), respectively.



The projection of the energy-momentum equation (\ref{eq:EM_con_01}) along
the four-velocity velocity $u^{\nu}$ expresses the conservation of energy
\cite{Rezzolla2013} and is given by
\begin{eqnarray}
\label{eq:EM_conser_03}
0 &  =  & u_{\nu}\partial_{\mu}T^{\mu\nu}\nonumber \\
  &  =  & u^{\alpha}\partial_{\alpha}e + (e + p- MB + B^2)\partial_{\alpha} u^{\alpha} 
  + Bu^{\alpha}\partial_{\alpha}B  \nonumber \\
  &  & + (MB -B^{2})u_{\mu}b^{\nu}\partial_{\nu}b^{\mu} \nonumber \\
  &  =  & u^{\alpha}\partial_{\alpha}e + (e + p)\partial_{\alpha}
u^{\alpha} + Mu^{\alpha}\partial_{\alpha}B\,,
\end{eqnarray}
where we have used that $u_{\alpha} b^{\alpha} = 0$ and Maxwell's
equations (\ref{eq:Maxwell_03}). Using the thermodynamical relations
(\ref{eq:Gibbs_03}), it is straightforward to conclude that a flow
conserves entropy, \ie 
\begin{equation}
\label{eq:entropy_02}
\partial_{\mu}(su^{\mu}) = 0\,,
\end{equation}
thus confirming that the thermodynamical relations (\ref{eq:Gibbs_03})
are consistent with the energy-momentum tensor (\ref{eq:EMT_01})
\cite{Caldarelli:2008ze,Gedalin:PRE1995,Huang:2009ue}.

Proceeding in a similar manner, the projection of the conservation
equation (\ref{eq:EM_con_01}) in the direction orthogonal to the
four-velocity $u^{\mu}$ expresses the conservation of momentum
\cite{Rezzolla2013} and is given by
\begin{eqnarray}
\label{eq:EM_conser_04}
0 & = & \Delta_{\nu\alpha}\partial_{\mu}T^{\mu\nu} \nonumber \\
  & = & (e + p-MB +
  B^{2})u^{\mu}\partial_{\mu}u_{\alpha} \nonumber \\ 
&& - \Delta_{\nu\alpha}\partial^{\nu}(p-MB
  + \frac{1}{2}B^{2})  \nonumber \\ 
&& + \Delta_{\nu\alpha}\partial_{\mu}[(MB-B^{2})b^{\mu}b^{\nu}]\,.
\end{eqnarray}
Later on, we will show that a Bjorken flow with nonzero magnetization
obeys Eq.\ (\ref{eq:EM_conser_04}).

The set of equations presented so far needs to be closed by an EOS and
because we are here searching for analytic solutions, we have considered
two EOSs that are particularly simple. The first one has been adopted
already in paper I and is given by 
\begin{equation}
\label{eq:EOS_1}
e = \frac{1}{c_{s}^{2}} p \,.
\end{equation}
where the sound speed is $c_s=1/\sqrt{3}$ when the fluid is
ultrarelativistic \cite{Rezzolla2013}. 

Clearly, the EOS \eqref{eq:EOS_1} is independent of the degree of
magnetization or of the strength of the magnetic field, which are however
accounted in the second EOS we consider and that is the one for a
conformal fluid in a four-dimensional spacetime and in the presence of a
magnetic field \cite{Caldarelli:2008ze}
\begin{equation}
\label{eq:EOS_2}
e = \frac{1}{c_{s}^{2}} p -2MB = 3p -2MB\,.
\end{equation}
The EOS above can be obtained through a conformal transformation
\cite{Caldarelli:2008ze,Bali:2014kia}, or simply by setting to zero the
trace of the energy-momentum tensor, \ie $T_{\ \ \mu}^{\mu} = 0$, and
obviously reduces to the ultrarelativistic-fluid EOS in the case of 
zero magnetization. In Refs.\ \cite{Bali:2014kia, Bali:2013esa}, these
two EOSs are named respectively EOSs for the ``$B$-scheme'' and the
``$\Phi$-scheme'', since they correspond to a fixed $B$ or a fixed
magnetic flux $\Phi$ during a conformal (compression) transformation,
respectively.

Before concluding this section, we will introduce a very important
theorem in ideal MHD, the frozen-flux theorem. With the help of
Eq.\ (\ref{eq:Maxwell_02}) and entropy conservation
(\ref{eq:entropy_02}), we find
\begin{equation}
\Delta_{\mu\nu}u^{\alpha}\partial_{\alpha}\left(\frac{B^{\mu}}{s}\right) =
\frac{1}{s}B^{\mu}\partial_{\mu}u_{\nu}\,,\label{eq:frozen_flux_01}
\end{equation}
which is the covariant form of the frozen-flux theorem. In the local rest
frame it reduces to
\begin{equation}
\label{eq:frozen_flux_02}
\frac{\partial}{\partial t}\left(\frac{\boldsymbol{B}}{s}\right) =
\frac{\boldsymbol{B}}{s}\cdot\nabla\boldsymbol{v}\,.
\end{equation}
In more physical terms, the condition \eqref{eq:frozen_flux_02} implies
that magnetic fields will evolve with the degrees of freedom of the fluid
\cite{Giacomazzo:2005jy, Giacomazzo:2007ti, Rezzolla2013, Roy:2015kma}.

\subsection{Generalized Bjorken flow}

We next discuss the generalization of the Bjorken flow investigated in
paper I to the case with nonzero magnetization. Here too, for the sake of
finding an analytic solution, we consider a longitudinally
boost-invariant Bjorken flow \cite{Bjorken:1982qr,Roy:2015kma}, whose
four-velocity is
\begin{equation}
u^{\mu} = \left(\frac{t}{\tau},0\,,0\,,\frac{z}{\tau}\right),\label{eq:velocity}
\end{equation}
where $\tau \equiv \sqrt{t^{2}-z^{2}}$  is the proper time. After
adopting Milne coordinates, $x^{\mu} = \left(\tau,x,y,\eta\right)$ with 
\begin{equation}
\eta \equiv \frac{1}{2}\ln\left(\frac{t
  + z}{t-z}\right)\,,
\end{equation}
being the spacetime rapidity. Using such coordinates, the four-velocity
simplifies to $u^{\mu} = (1,\boldsymbol{0})$, with the directional
derivative and four-divergence given respectively by
\begin{equation}
u^{\mu}\partial_{\mu} = \partial_{\tau}\,, \qquad \qquad
\partial_{\alpha} u^{\alpha} = \frac{1}{\tau}\,.
\end{equation}

In relativistic heavy-ion collisions, the magnetic field points normally into
the transverse direction, \ie perpendicular to the $z$ direction if this
is taken to the beam axis. Hence, we choose the three-vector
$\boldsymbol{B}$ to be parallel to the $y$ direction and homogeneous in
the transverse plane, \ie
\begin{equation}
\boldsymbol{B} = B\boldsymbol{e}_{y}\,,\label{eq:magentic_01}
\end{equation}
In this case, the magnetic field does not change in Milne coordinates
and the frozen-flux theorem (\ref{eq:frozen_flux_01}) gives
\begin{equation}
\partial_{\tau}\left(\frac{B}{s}\right) = 0\,,
\end{equation}
which implies that $B/s$ is conserved, or more explicitly, that
\begin{equation}
\label{eq:frozen_flux_00}
\frac{B}{B_{0}} = \frac{s}{s_{0}}\,,
\end{equation}
with $B_{0}$ and $s_{0}$ being the initial magnetic field and the entropy
density, respectively. As already mentioned above, the condition
\eqref{eq:frozen_flux_00} means that the magnetic field is advected and
distorted by the fluid motion in the same way as a fluid element
\cite{Roy:2015kma}; identical considerations apply also for astrophysical
plasmas, where the rest-mass density is normally used in place of the
entropy density \cite{Giacomazzo:2005jy, Giacomazzo:2007ti,
  Rezzolla2013}. Inserting the velocity from Eq.\ (\ref{eq:velocity})
into Eq.\ (\ref{eq:entropy_02}), Eq.\ (\ref{eq:frozen_flux_00}) becomes
\begin{equation}
\label{eq:frozen_flux_03}
\frac{B}{B_{0}} = \frac{s}{s_{0}} =
\frac{\tau_{0}}{\tau}\,.
\end{equation}
thus introducing a simple scaling with proper time.


\begin{figure*}
\begin{centering}
\includegraphics[width=0.99\columnwidth]{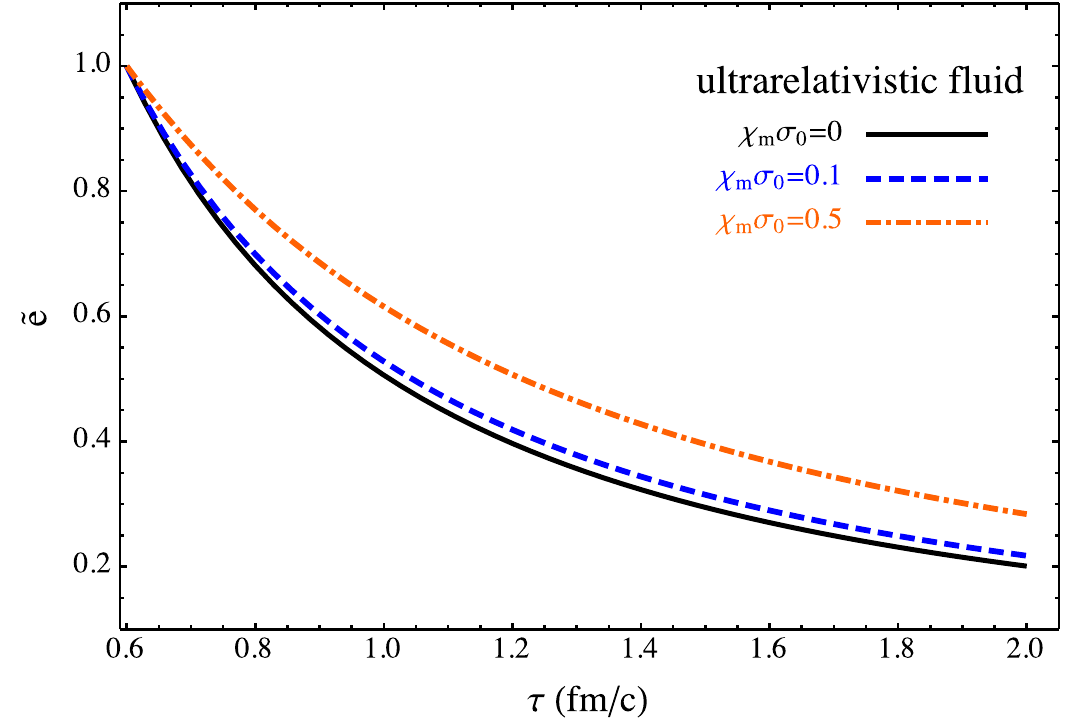} 
\hskip 0.5cm
\includegraphics[width=0.99\columnwidth]{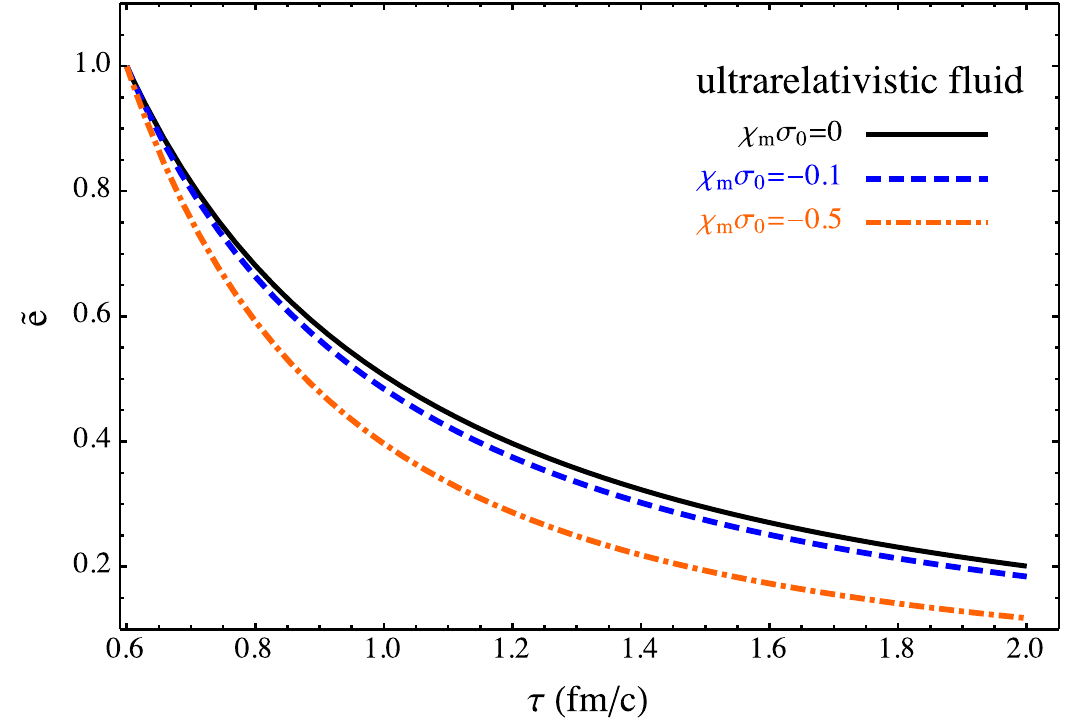} 
\caption{The evolution of the energy density, Eq.\ (\ref{eq:sol_01}),
  with $c_{s}^{2} = 1/3$ and the EOS (\ref{eq:EOS_1}). In the left
  panel, the solid, dashed, and dash-dotted lines are for
  $\chi_{m}\sigma_{0} = 0,0.1,0.5$, respectively, while in the right
  panel, the solid, dashed, and dash-dotted lines are for
  $\chi_{m}\sigma_{0} = 0,-0.1,-0.3$, respectively. \label{fig:sol_01}}
\end{centering}
\end{figure*}

Finally, let us consider the momentum conservation
Eqs.\ (\ref{eq:EM_conser_04}), which will ultimately yield the well-known
Bjorken scaling law. In the case in which the susceptibility $\chi_{m}$
is a constant, the last term in Eq.\ (\ref{eq:EM_conser_04}) reduces to
\begin{eqnarray}
&&\Delta_{\nu}^\alpha\partial_{\mu}[(MB-B^{2})b^{\mu}b^{\nu}] =
  \nonumber\\ 
&& \hskip 0.75cm
  (\chi_{m}-1)[\partial_{\mu}(B^{\mu}B^{\alpha}) -
    u_{\nu}u^{\alpha}\partial_{\mu}(B^{\mu}B^{\nu})]\,,
\end{eqnarray}
Using now the four-velocity (\ref{eq:velocity}) and the magnetic field as
given by Eqs.\ (\ref{eq:magentic_01}), (\ref{eq:frozen_flux_03}), it is
straightforward to show that such a term vanishes, reducing
Eq.\ (\ref{eq:EM_conser_04}) to 
\begin{equation}
\label{eq:BF_gen}
0 = (e + p-MB +
B^{2})\partial_{\tau}u_{\mu}-\Delta_{\nu\mu}\partial^{\nu}(p-MB +
\frac{1}{2}B^{2})\,.
\end{equation}
When $\mu = \eta$, Eq.\ \eqref{eq:BF_gen} reads

\begin{equation}
\partial_{\eta}\left(p-MB + \frac{1}{2}B^{2}\right) = 0\,,
\end{equation}
thus showing that all thermodynamic variables depend only on the proper time
$\tau$, which is the time-honored Bjorken scaling. On the other hand,
when $\mu = x, y$, Eq.\ \eqref{eq:BF_gen} becomes
\begin{equation}
\partial_{\tau}u_{i}-\frac{1}{e + p-MB + B^{2}}\partial_{i}
\left(p-MB +
\frac{1}{2}B^{2}\right) = 0\,.
\end{equation}
Not surprisingly, when the pressure and the magnetic fields are uniform,
the second term in the equation above vanishes, implying that the motion
will be geodetic, \ie with constant velocity.

\section{Energy-density evolution}
\label{sec:Energy_ideal}

\subsection{Ultrarelativistic fluid}

Using Eq.\ (\ref{eq:EM_conser_03}) and the definition of the
susceptibility in Eq.\ (\ref{eq:susceptibility_01}), the energy
conservation equation \eqref{eq:EM_conser_03} reads 
\begin{equation}
\label{eq:sol_eq_01}
\partial_{\tau} e  + \frac{e + p}{\tau} +
\frac{1}{2}\chi_{m} \partial_{\tau} B^{2} =
0\,.
\end{equation}
Before we seek for solutions, let us remark that if $\chi_{m} = 0$ and
thus $M = 0$, \ie if we neglect the magnetization of the fluid, then
Eq.\ (\ref{eq:sol_eq_01}) is the same as in the standard Bjorken flow
without magnetic fields. As discussed in paper I, we would still have the
contribution of the magnetic field in the energy-momentum tensor
(\ref{eq:EMT_01}), but this would not affect the decay of the energy
density \cite{Roy:2015kma}.

With the help of the frozen-flux theorem (\ref{eq:frozen_flux_03}),
Eq.\ (\ref{eq:EMT_01}) can be rewritten as
\begin{equation}
\partial_{\tau} e + \frac{e +
  p}{\tau}-\frac{\chi_{m}B_{0}^{2}}{\tau^{3}}\tau_{0}^{2} =
0\,.\label{eq:sol_eq_02}
\end{equation}
so that, after introducing the dimensionless quantities
\begin{equation}
\tilde{e} \equiv \frac{e}{e_{0}}\,,
\qquad \qquad 
\sigma_{0} \equiv \frac{B_{0}^{2}}{e_{0}}\,,
\end{equation}
and using the EOS \eqref{eq:EOS_1} for an ultrarelativistic fluid,
Eq.\ (\ref{eq:sol_eq_02}) can be written as 
\begin{equation}
\partial_{\tau} \tilde{e} + (1 +
c_{s}^{2})\frac{\tilde{e}}{\tau}-\frac{\chi_{m}\sigma_{0}\tau_{0}^{2}}{\tau^{3}}
= 0\,.\label{eq:sol_eq_03}
\end{equation}

Setting as initial condition $\tilde{e}_{0} \equiv \tilde{e}(\tau_{0}) =
1$, the solution of this differential equation is given by 
\begin{equation}
\label{eq:sol_01}
\tilde{e}(\tau) = \left(\frac{\tau_{0}}{\tau}\right)^{1 +
  c_{s}^{2}}-\frac{\chi_{m}\sigma_{0}}{1 -
  c_{s}^{2}}\left[\left(\frac{\tau_{0}}{\tau}\right)^{2} -
  \left(\frac{\tau_{0}}{\tau}\right)^{1 +
    c_{s}^{2}}\right]\,.
\end{equation}

Recalling now that the total energy density is given by the double
contraction of the energy-momentum tensor along the four-velocity
\begin{equation}
e_{\rm tot} \equiv T^{\mu\nu}u_{\mu}u_{\nu} = e + \frac{1}{2}B^{2}\,
\end{equation}
and after introducing the dimensionless total energy density as
\begin{equation}
\label{eq:sol_02}
\tilde{e}_{\rm tot}  \equiv  \frac{e_{\rm tot}}{e_{0}} = \tilde{e} +
\frac{1}{2}\sigma_{0}\left(\frac{B}{B_{0}}\right)^{2} \,,
\end{equation}
the analytic evolution of the energy density in a Bjorken flow with
nonzero magnetic susceptibility is given by
\begin{eqnarray}
\label{eq:tot_eng_01}
\tilde{e}_{\rm tot} & = & 
\left(\frac{\tau_{0}}{\tau}\right)^{1 + c_{s}^{2}}  \\ 
& \phantom{=} & + \sigma_{0}\left\{
\frac{1}{2}\left(\frac{\tau_{0}}{\tau}\right)^{2} -
\frac{\chi_{m}}{1-c_{s}^{2}}\left[\left(\frac{\tau_{0}}{\tau}\right)^{2}
  - \left(\frac{\tau_{0}}{\tau}\right)^{1 +
    c_{s}^{2}}\right]\right\}\,, \nonumber 
\end{eqnarray}
showing that, at late times, the energy density decays as $\sim1/\tau^{1 +
  c_{s}^{2}}$.

To better understand the properties of the solution \eqref{eq:tot_eng_01}
we can take a closer look at the differential equation
(\ref{eq:sol_eq_01}), which tells us that there are two sources to the
variation of the energy density. The first one is proportional to $-(e +
p)\partial_{\alpha} u^{\alpha} = -(e + p)/\tau$ and is related to the
expansion of the fluid. It obviously leads to an adiabatic decrease of
the energy density. The second term is proportional to
$-M\partial_{\tau}B\propto\chi_{m}/\tau^{3}$ and will have a different
behaviour depending on the magnetic properties of the fluid. In
particular, if the fluid is paramagnetic (\ie if $\chi_{m}>0$), then the
fluid will gain energy from the magnetic field and the rate of energy-density 
decrease will be smaller than without magnetization. On the other
hand, if the fluid is diamagnetic (\ie if $\chi_{m}<0$), then the fluid
has to spend additional energy to expel the magnetic field, leading to an
energy-density decrease that is much faster than without magnetization. This
behaviour is summarized in Fig.\ \ref{fig:sol_01}, where we plot the
solution (\ref{eq:sol_01}) for $\chi_{m}\sigma_{0} = 0,\pm0.1,\pm0.5$
in the case $c_{s}^{2} = 1/3$. The left panel shows the evolution of the
dimensionless energy density relative to a fluid with positive
susceptibility ($\chi_{m} \sigma_0 =0.1, 0.5$), while the right one
refers to a fluid with negative susceptibility ($\chi_{m} \sigma_0
=-0.1, -0.5$). In both cases, the black lines indicate the case with zero
susceptibility, \ie the classical Bjorken evolution.

\subsection{Magnetized conformal fluid}

Next, we turn to discuss the solutions for a fluid with the EOS
(\ref{eq:EOS_2}) for a magnetized conformal fluid. In this case,
Eqs.\ (\ref{eq:sol_eq_01}) and (\ref{eq:sol_eq_02}) become
\begin{equation}
\label{eq:sol_eq_04}
\partial_{\tau} \tilde{e} +
\frac{4}{3}\frac{\tilde{e}}{\tau} -
\frac{1}{3}\frac{\chi_{m}\sigma_{0}}{\tau^{3}} = 0\,,
\end{equation}
whose solution can be found as in the previous case and reads
\begin{equation}
\label{eq:sol_01b}
\tilde{e}(\tau) = \left(\frac{\tau_{0}}{\tau}\right)^{4/3} -
\frac{1}{2}\chi_{m}\sigma_{0}\left[\left(\frac{\tau_{0}}{\tau}\right)^{2}
  - \left(\frac{\tau_{0}}{\tau}\right)^{4/3}\right]\,,
\end{equation}
while for the total dimensionless energy density it is given by
\begin{equation}
\tilde{e}_{\rm tot} = 
\left(\frac{\tau_{0}}{\tau}\right)^{4/3} + \frac{\sigma_{0}}{2}\left\{
\left(\frac{\tau_{0}}{\tau}\right)^{2} -
\chi_{m}\left[\left(\frac{\tau_{0}}{\tau}\right)^{2} -
  \left(\frac{\tau_{0}}{\tau}\right)^{4/3}\right]\right\}\,.
\end{equation}

\begin{figure}
\begin{centering}
\includegraphics[width=0.99\columnwidth]{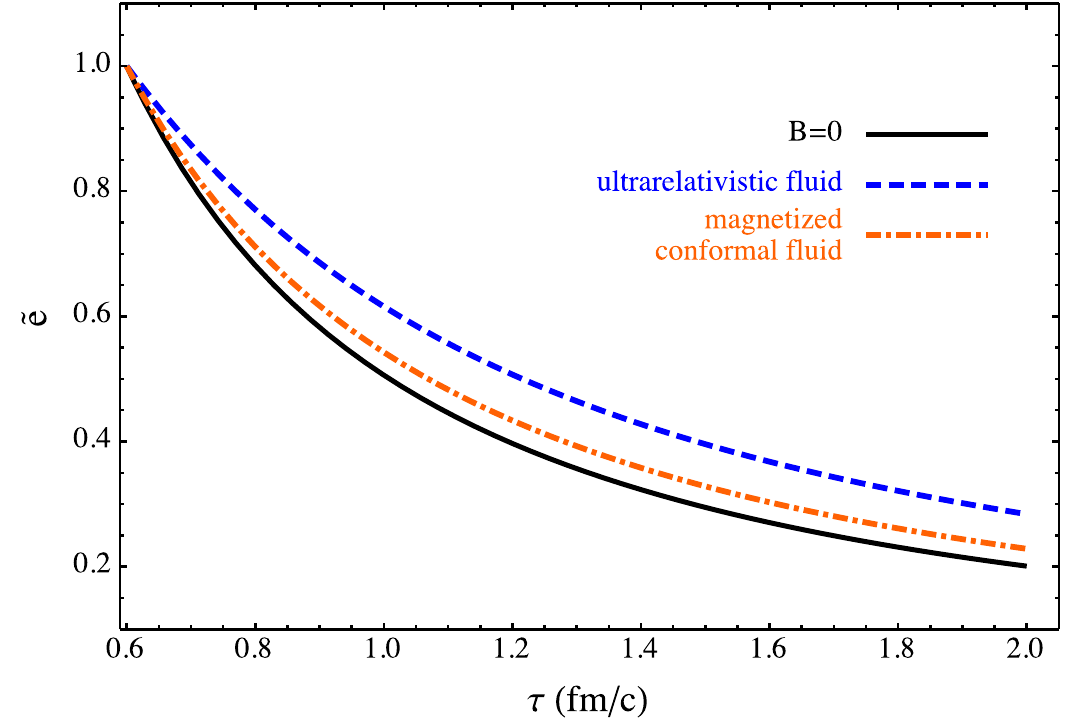}
\par\end{centering}
\caption{The evolution of the energy density. The black line is the
  solution without magnetic fields. The dashed and dash-dotted lines are
  for the solutions for the EOSs considered with $\chi_{m}\sigma_{0} =
  0.5$, respectively. \label{fig:sol_02}}
\end{figure}

Comparing the solution (\ref{eq:sol_02}) with \eqref{eq:sol_01b}, it is
easy to conclude that the solution of a magnetized conformal fluid is
the same one as in the case of an ultrarelativistic fluid, which however
has smaller effective magnetic susceptibility, \ie $\chi_m \to
\chi_{m}/3$. In Fig.\ \ref{fig:sol_02}, we compare the solutions
(\ref{eq:sol_02}) and \eqref{eq:sol_01b} for the two different EOSs
considered. Note that for a positive susceptibility, the fluid with the
EOS \eqref{eq:EOS_2} of a magnetized conformal fluid will gain less
energy from the magnetic field than for the EOS \eqref{eq:EOS_1} of an
ultrarelativistic fluid, thus being closer to the case when the magnetic
field is absent. One can draw similar conclusions for negative
susceptibilities.

\section{Energy-density evolution with an external magnetic field}
\label{sec:power law}

\subsection{Ultrarelativistic fluid}

We now extend our discussion to case of an external magnetic field which
is suitably tuned to decay following a power law in proper time, \ie
\begin{equation}
\label{eq:power_law_01}
B(\tau) = B_{0}\left(\frac{\tau_{0}}{\tau}\right)^{a}\,,
\end{equation}
where $a$ is a constant and the case $a = 1$ falls back to the case
discussed in the previous section [\cf
  Eq.\ \eqref{eq:frozen_flux_03}]. Given the typical and very short
timescales involved in heavy-ion collisions, this scenario is rather
unrealistic, but we consider it here partly because it has been
investigated also in paper I and partly because it allows us to obtain
another interesting analytic solution. Another important simplifying
assumption is that we will take the external field to be much stronger
than any magnetic field produced at the collision. This implies that we
can neglect the latter and, more importantly, that such external field
does not have to satisfy Maxwell's equations (\ref{eq:Maxwell_02})
coupled to the fluid. 

Under these somewhat academic assumptions, the
energy-momentum conservation Eq.\ (\ref{eq:EM_con_03}) with the Bjorken
velocity becomes
\begin{eqnarray}
\label{eq:EM_con_03}
\partial_{\tau}e + \frac{e + p-MB + B^{2}}{\tau} +
\frac{1}{2}\partial_{\tau}B^{2} = 0\,.
\end{eqnarray}
where we used the fact that $b^{\mu}$ is non-vanishing only in the $y$
direction, so that $-B^{2}u_{\mu}b^{\nu}\partial_{\nu}b^{\mu} = 0=
MBu_{\mu}b^{\nu}\partial_{\nu}b^{\mu}$. Inserting
Eq.\ (\ref{eq:power_law_01}) into Eq.\ (\ref{eq:EM_con_03}) yields
\begin{equation}
\partial_{\tau} e  + \frac{e + p}{\tau} +
(1-a-\chi_{m})\frac{B_{0}^{2}\tau_{0}^{2a}}{\tau^{2a + 1}} =
0\,.\label{eq:sol_eq_5}
\end{equation}
It is simple to check that when $a = 1$, Eq.\ (\ref{eq:sol_eq_5}) reduces
to Eq.\ (\ref{eq:sol_eq_03}) and when $\chi_{m} = 0$,
Eq.\ (\ref{eq:sol_eq_5}) is also consistent with the results of
Ref.\ \cite{Roy:2015kma}.

Using the EOS (\ref{eq:EOS_1}), we find 
\begin{equation}
\partial_{\tau} \tilde{e} + (1 +
c_{s}^{2})\frac{\tilde{e}}{\tau} + (1-a-\chi_{m})\sigma_{0}
\frac{\tau_{0}^{2a}}{\tau^{2a + 1}} = 0\,.
\end{equation}
The solution is 
\begin{equation}
\tilde{e}(\tau) = \left(\frac{\tau_{0}}{\tau}\right)^{1 + c_{s}^{2}} -
\sigma_{0}\frac{1 - a - \chi_{m}}{1 + c_{s}^{2} -
  2a}\left[\left(\frac{\tau_{0}}{\tau}\right)^{2a} -
  \left(\frac{\tau_{0}}{\tau}\right)^{1 +
    c_{s}^{2}}\right]\label{eq:sol_05_a}
\end{equation}
for $a\neq(1 + c_{s}^{2})/2$, and
\begin{equation}
\label{eq:sol_05_b}
\tilde{e}(\tau) = \left(\frac{\tau_{0}}{\tau}\right)^{1 + c_{s}^{2}} +
\frac{1}{2}\sigma_{0}(1-c_{s}^{2}-2\chi_{m})
\left(\frac{\tau_{0}}{\tau}\right)^{1 +
  c_{s}^{2}}\log\left(\frac{\tau_{0}}{\tau}\right)\,
\end{equation}
for $a = (1 + c_{s}^{2})/2$. One can also get Eq.\ (\ref{eq:sol_05_b}) by
taking the limit $a\rightarrow(1 + c_{s}^{2})/2$ for the solution
(\ref{eq:sol_05_a}). The normalized total energy density is then
\begin{eqnarray}
&& \hskip -0.4cm \tilde{e}_{\rm tot} =  
\left(\frac{\tau_{0}}{\tau}\right)^{1 +  c_{s}^{2}}
\nonumber \\
&&
- \sigma_{0}\frac{1-a-\chi_{m}}{1 + c_{s}^{2}-2a}
\left[\left(\frac{\tau_{0}}{\tau}\right)^{2a} - 
\left(\frac{\tau_{0}}{\tau}\right)^{1
    + c_{s}^{2}}\right] +
\frac{1}{2}\sigma_{0}\left(\frac{\tau_{0}}{\tau}\right)^{2a}\,,
\nonumber \\
\end{eqnarray}
for $a\neq(1 + c_{s}^{2})/2$, and 
\begin{eqnarray}
&& \hskip -0.4cm \tilde{e}_{\rm tot} =  
\left(\frac{\tau_{0}}{\tau}\right)^{1 + c_{s}^{2}} +
\frac{1}{2}\sigma_{0}(1-c_{s}^{2}-2\chi_{m})
\left(\frac{\tau_{0}}{\tau}\right)^{1 +
  c_{s}^{2}}\log\left(\frac{\tau_{0}}{\tau}\right) 
\nonumber \\
&&
+\frac{1}{2}\sigma_{0} \left(\frac{\tau_{0}}{\tau}\right)^{2a}\,,
\end{eqnarray}
for $a = (1 + c_{s}^{2})/2$.

\begin{figure*}
\begin{centering}
\includegraphics[width=0.99\columnwidth]{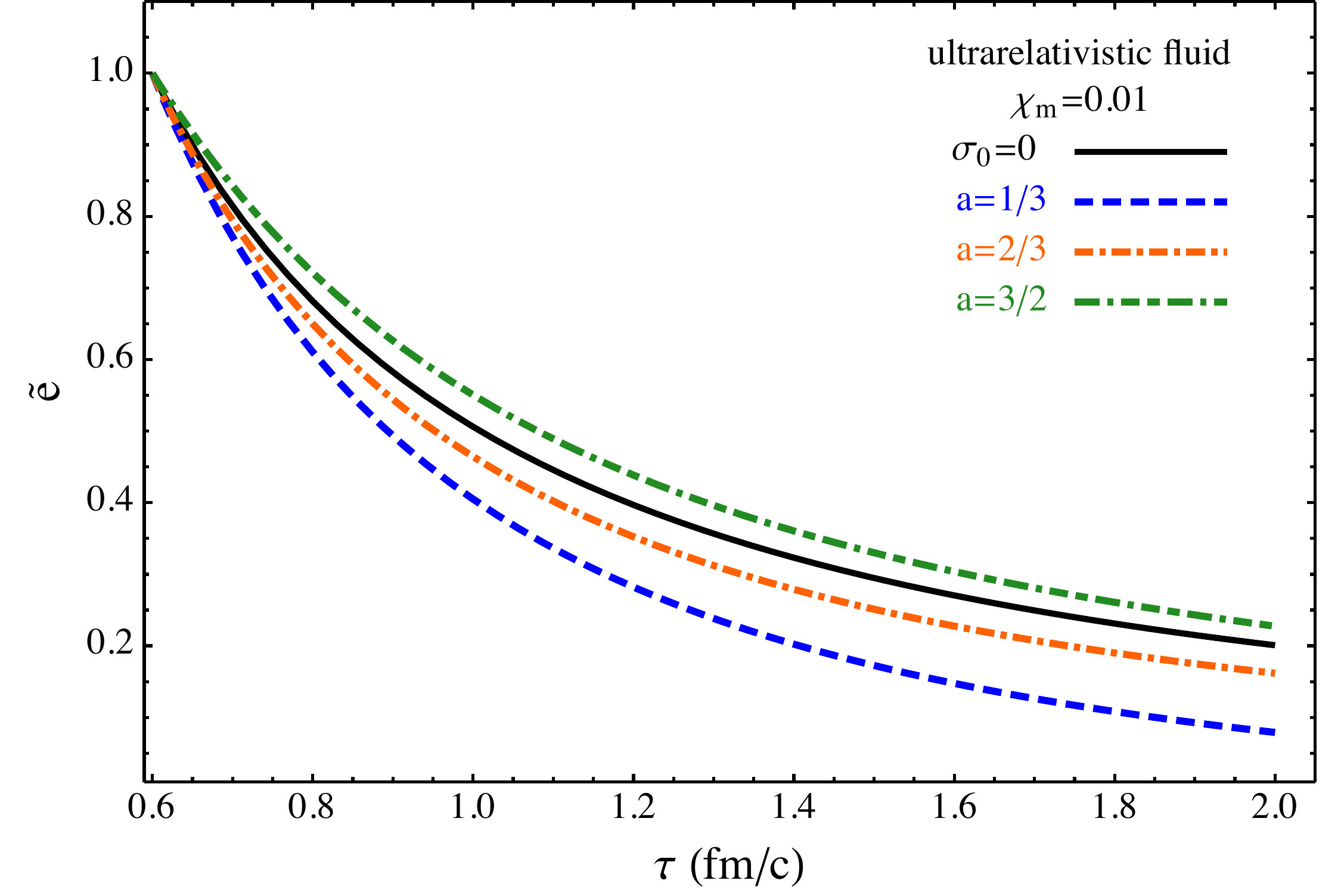}
\hskip 0.5cm
\includegraphics[width=0.99\columnwidth]{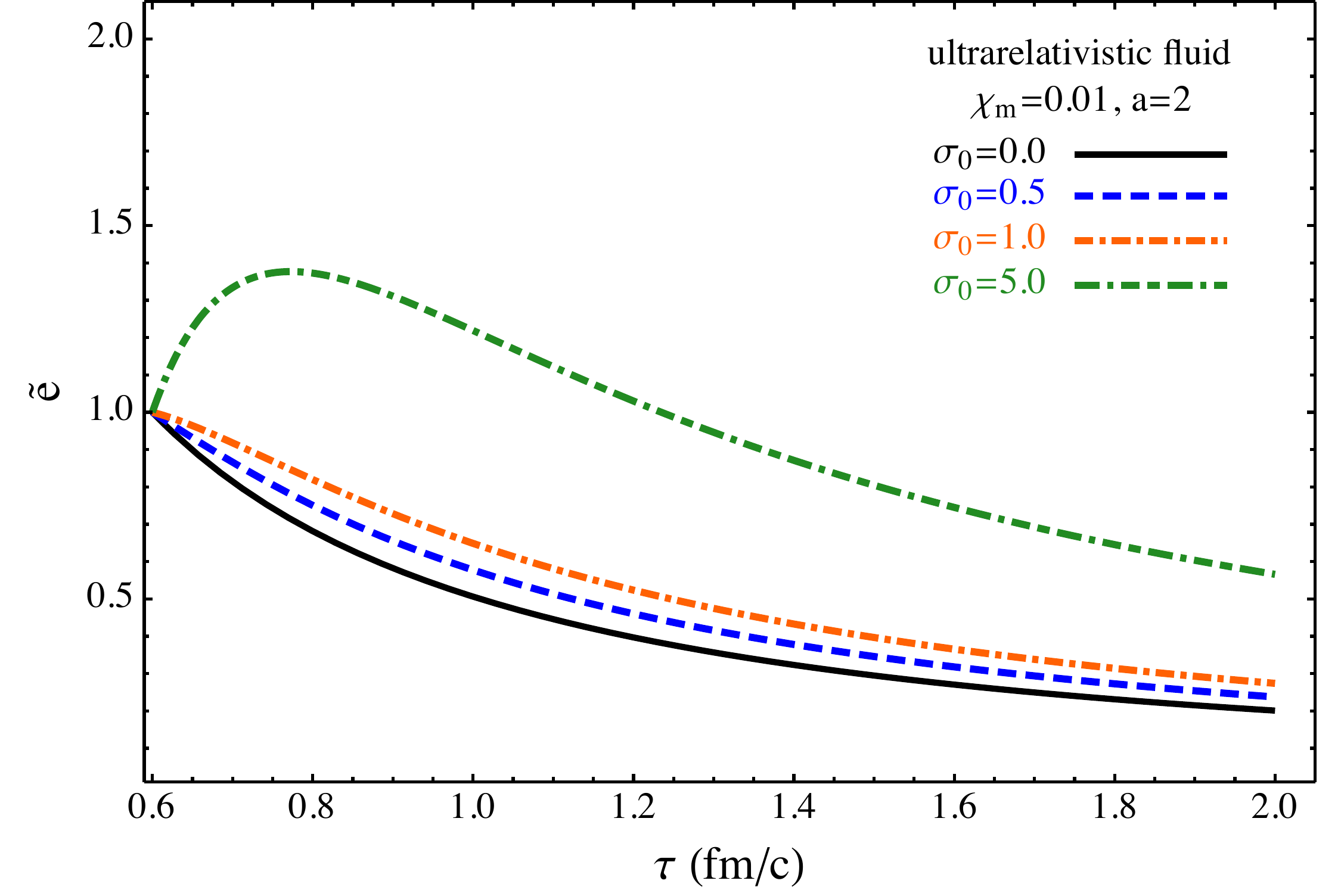}
\caption{{Left panel}: the solutions (\ref{eq:sol_05_a}),
  (\ref{eq:sol_05_b}) for different values of the parameters. We choose
  $\chi_{m} = 0.01$. The black solid line is for the case without
  magnetic field. For $\sigma_{0} = 0.5$, the blue dashed, orange
  dash-dotted, and green dash-dotted lines are for $a = 1/3,2/3,3/2$,
  respectively. {Right panel}: the same as in the left panel but
  for $\chi_{m} = 0.01$ and $a = 2$. The black solid line is for the case
  without magnetic fields. The blue dashed, orange dash-dotted, and
  green dash-dotted lines are for $\sigma_{0} = 0.5,1.0,5.0$,
  respectively.}
\label{fig:03}
\end{centering}
\end{figure*}

Since the term $\left[\left({\tau_{0}}/{\tau}\right)^{2a} -
  \left({\tau_{0}}/{\tau}\right)^{1 + c_{s}^{2}}\right]/(1 +
c_{s}^{2}-2a)$ is always positive, from Eq.\ (\ref{eq:sol_05_a}) we see
that the effect of nonzero magnetization enters only through the
prefactor $1-a-\chi_{m}$. More precisely, when $a>1-\chi_{m}$,
$\tilde{e}$ decays more slowly than in the case without magnetic fields,
and when $0<a<1-\chi_{m}$, $\tilde{e}$ decays faster. One can reach the
same conclusion by analyzing the sign of the last term in
Eq.\ (\ref{eq:sol_eq_5}). We can demonstrate our conclusion in two limits
and, for simplicity, we assume $\chi_{m}>0$. For $a\rightarrow0$, the
magnetic field is constant in proper time and does not evolve with the
fluid. Thus, the fluid energy density must decay faster in order to
sustain this constant magnetic field. For $a\rightarrow\infty$, the
magnetic field decays very rapidly and its energy will be transferred to
the fluid due to the energy-conservation law. So, one can expect a peak
of the energy density near the initial time, which is associated with a
resistive ``reheating'' of the fluid \cite{Roy:2015kma}.

We show these solutions in Fig.\ \ref{fig:03}. Since $\chi_{m}$ is very
small for the QGP, we choose a typical value $\chi_{m} = 0.01$ in the
left panel of Fig.\ \ref{fig:03}. The black solid line is for the case
without magnetic field, $\sigma_0 = 0$, while the blue dashed, orange
dash-dotted, and green dash-dotted lines are for $a = 1/3,2/3,3/2$,
respectively, for the case $\sigma_{0} = 0.5$. For $a>1-\chi_{m}$ the
energy density decays more slowly and for $a<1-\chi_{m}$, it decays
faster than without magnetization, respectively. In the right panel of
Fig.\ \ref{fig:03}, fixing $\chi_{m} = 0.01$ and $a = 2$, the blue
dashed, orange dash-dotted, and green dash-dotted lines are for
$\sigma_{0} = 0.5,1.0,5.0$, respectively. If $\sigma_{0}$ is large
enough, we can observe the initial ``reheating'' effect.

\subsection{Magnetized conformal fluid}

We conclude this section by considering the evolution under external
magnetic field \eqref{eq:power_law_01} when the fluid obeys the EOS
(\ref{eq:EOS_2}) for a magnetized conformal fluid. In this case,
Eq.\ (\ref{eq:sol_eq_5}) becomes
\begin{equation}
\partial_{\tau} \tilde{e} +
\frac{4}{3}\frac{\tilde{e}}{\tau} +
\sigma_{0}\left(1-a-\frac{1}{3}\chi_{m}\right)\frac{\tau_{0}^{2a}}{\tau^{a
    + 1}} = 0\,.\label{eq:sol_eq_06}
\end{equation}
The solution can be obtained similarly as above and reads
\begin{equation}
\tilde{e}(\tau) = \left(\frac{\tau_{0}}{\tau}\right)^{4/3} -
\frac{\sigma_{0}}{2}\frac{3 - 3a - \chi_{m}}{2 - 3a}
\left[\left(\frac{\tau_{0}}{\tau}\right)^{2a} -
  \left(\frac{\tau_{0}}{\tau}\right)^{4/3}\right] \label{eq:sol_06a}
\end{equation}
for $a\neq2/3$, and
\begin{equation}
\tilde{e}(\tau) = \left(\frac{\tau_{0}}{\tau}\right)^{4/3} +
\frac{\sigma_{0}}{3}(1 - \chi_{m})
\left(\frac{\tau_{0}}{\tau}\right)^{4/3}
\log\left(\frac{\tau_{0}}{\tau}\right)\label{eq:sol_06b}
\end{equation}
for $a = 2/3$. The normalized total energy density is
\begin{eqnarray}
&& \hskip -0.4 cm \tilde{e}_{\rm tot} =
\left(\frac{\tau_{0}}{\tau}\right)^{4/3}  
\nonumber \\
&& \hskip 0.0cm
+ \frac{\sigma_{0}}{2}\left\{
\left(\frac{\tau_{0}}{\tau}\right)^{2} - \frac{3 - 3a - \chi_{m}}{2 -
  3a}\left[\left(\frac{\tau_{0}}{\tau}\right)^{2a} -
  \left(\frac{\tau_{0}}{\tau}\right)^{4/3}\right]\right\}
\nonumber \\
\end{eqnarray}
for $a\neq2/3$, and 
\begin{equation}
\tilde{e}_{\rm tot} = \left(\frac{\tau_{0}}{\tau}\right)^{4/3} +
\frac{\sigma_{0}}{3}(1-\chi_{m})
\left(\frac{\tau_{0}}{\tau}\right)^{4/3}\log\left(\frac{\tau_{0}}{\tau}\right)
+ \frac{\sigma_{0}}{2} \left(\frac{\tau_{0}}{\tau}\right)^{2}
\end{equation}
for $a=2/3$.
The behaviour of Eq.\ (\ref{eq:sol_06a}) can be discussed in a manner
similar to the previous case of an ultrarelativistic fluid. When
$a>1-\chi_{m}/3$, $\tilde{e}$ will decay more slowly than in the case
without magnetic field, and when $0<a\leq1-\chi_{m}/3$, $\tilde{e}$ will
decay faster than in the case without magnetic field.

\begin{figure*}
\begin{centering}
\includegraphics[width=0.99\columnwidth]{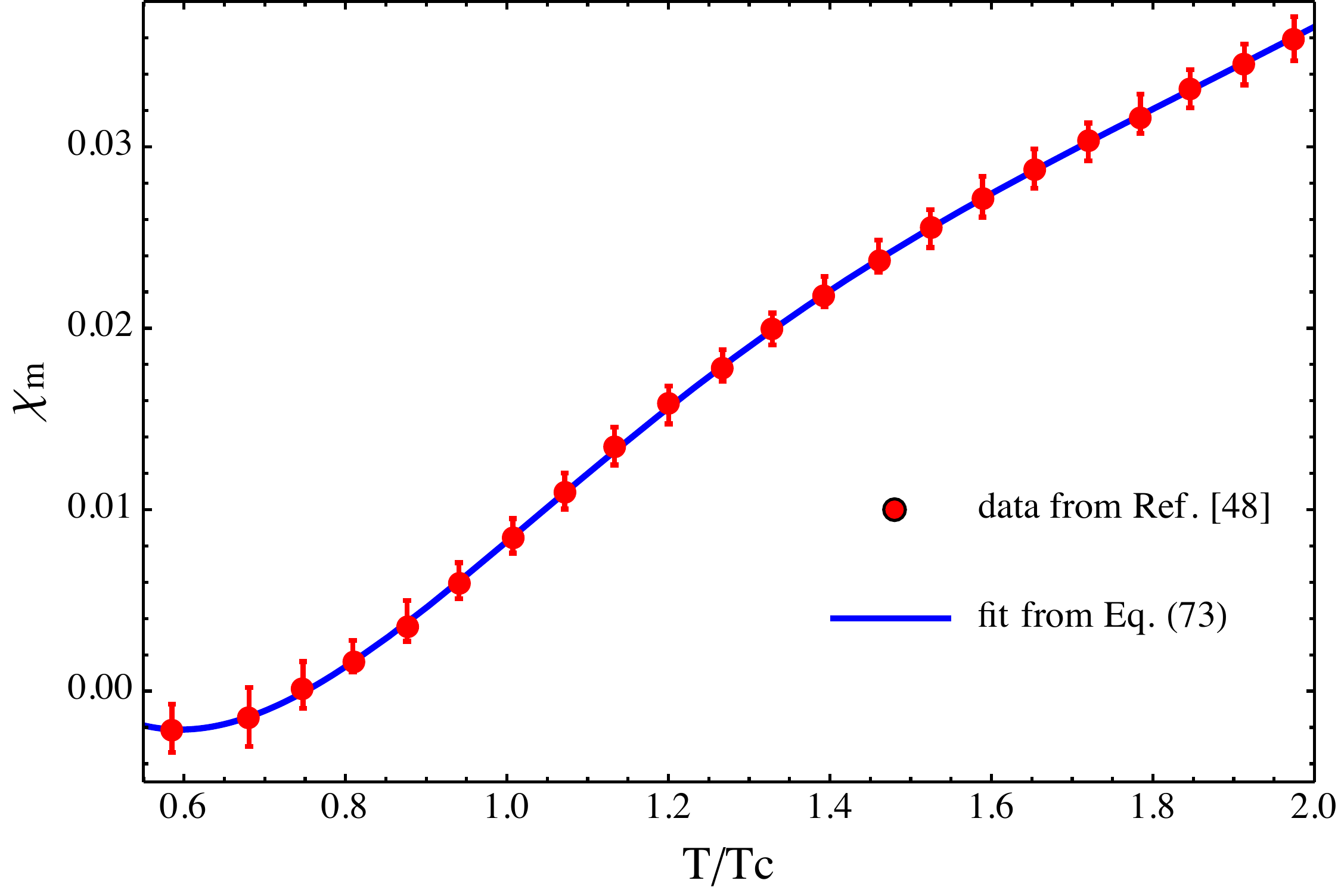}
\hskip 0.5cm
\includegraphics[width=0.99\columnwidth]{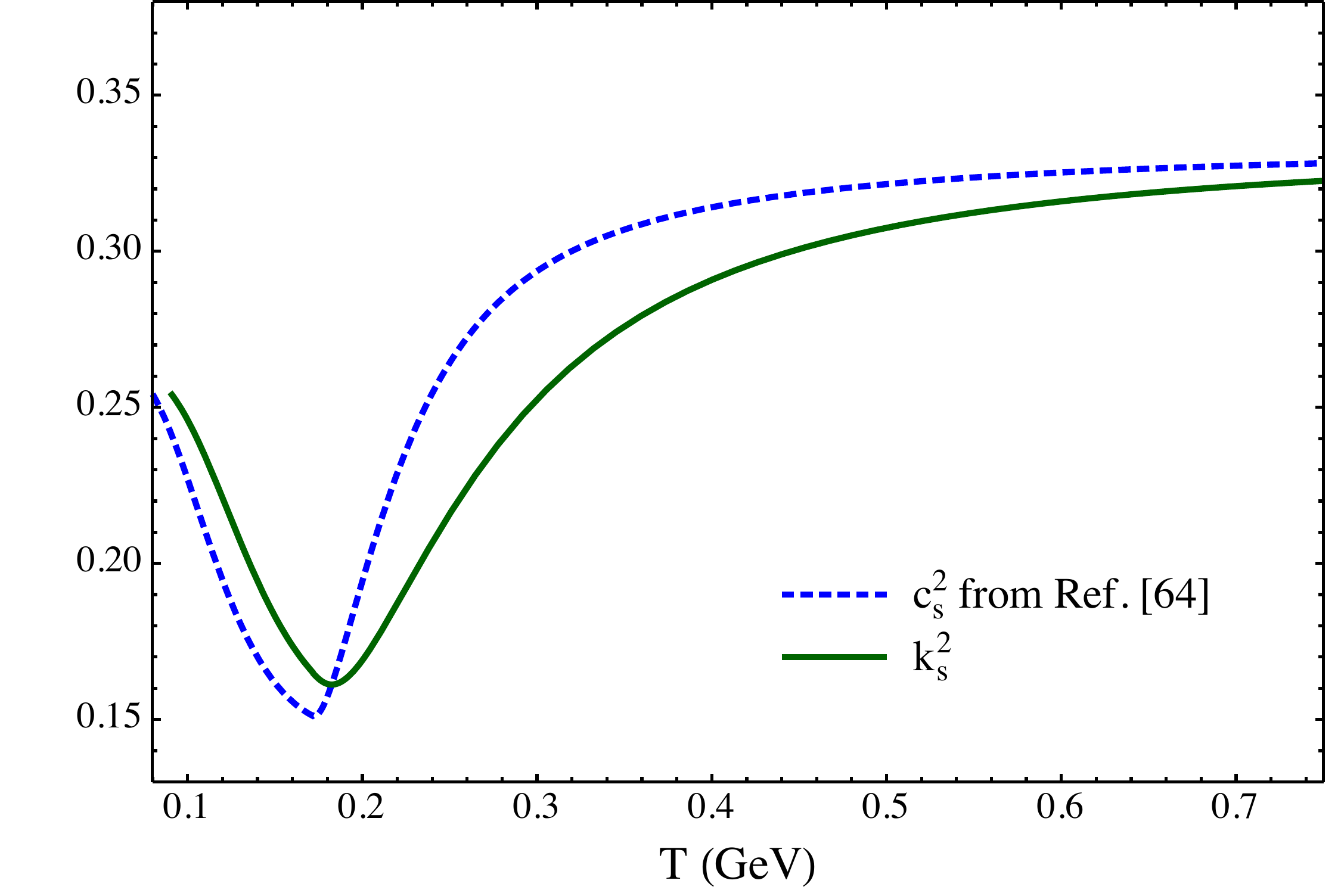}
\caption{Left panel: magnetic susceptibility $\chi_{m}$;
the filled red circles refer to the data in Ref.\ \cite{Bali:2014kia}, 
while the solid line shows the fit given by Eq.\ (\ref{eq:chi_T_fit}). 
Right panel: behaviour of
$k_{s}^{2}=p/e$ and of  $c_s^2$. The dashed line shows the speed of sound
 given by Eq.\ (\ref{eq:cs_T_1}) from the parametrization of 
Ref.\  \cite{Huovinen:2009yb}. 
Finally, the solid green line provides the ratio between the 
pressure and energy density;
 note the close similarity between $k_s^2$ and $c_s^2$.
\label{fig:Parametrization}}
\end{centering}
\end{figure*}

\section{Energy-density evolution with temperature-dependent 
magnetic susceptibility}
\label{sec:Temperature-dependent-magnetic}

\begin{figure*}
\centering{}\includegraphics[width=0.99\columnwidth]{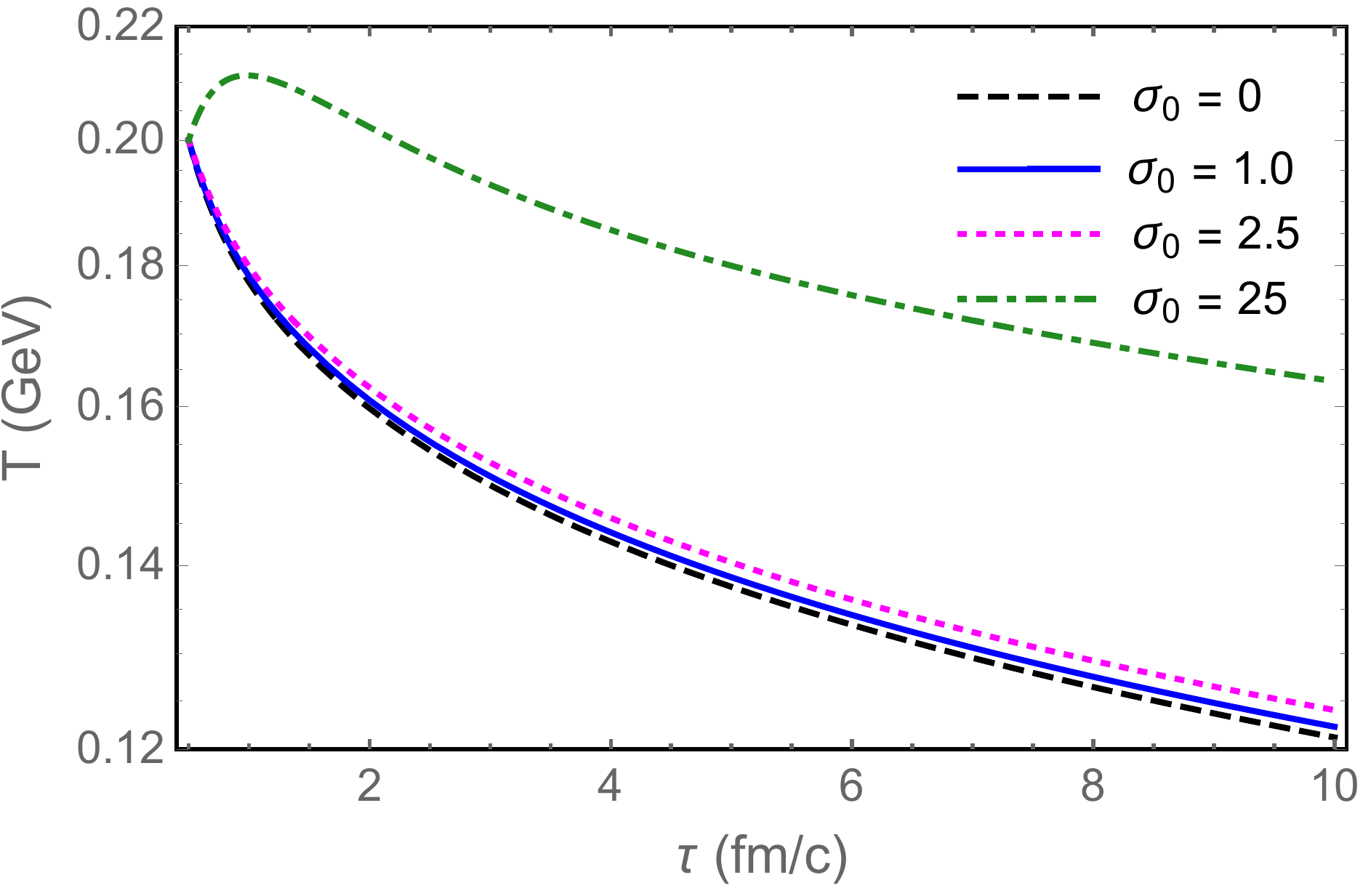} \hskip
0.5cm
\includegraphics[width=0.99\columnwidth]{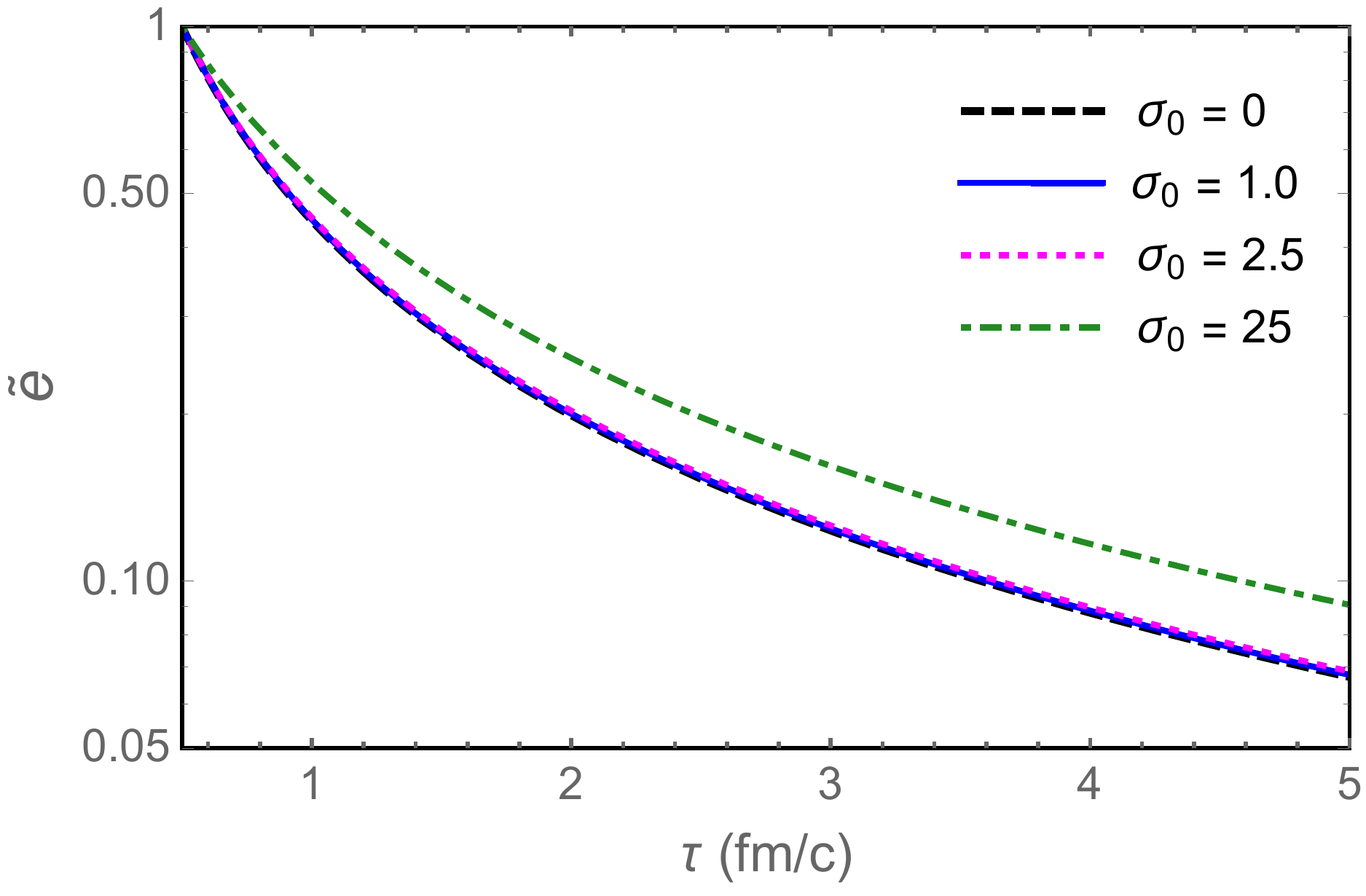} \caption{Evolution of
  temperature $T$ (left panel) and normalized energy density
  $\tilde{e}=e/e_{0}$ (right panel) as a functions of proper time
  $\tau$. We choose initial temperature $T_{0}=200$ MeV, initial time
  $\tau_{0}=0.5$ fm/c, and critical temperature $T_{c}=180$ MeV. The
  black dashed, blue solid, magenta dotted, and green dot-dashed lines
  are for $\sigma_{0}=0,1.0,2.5,25$,
  respectively. \label{fig:Evolution_02}}
\end{figure*}

After the rather academic discussion of the previous section, we will now
consider a more realistic scenario. For the sake of simplicity, we will
choose the temperature $T$ as independent variable and express all
quantities as a function of $T$. We also choose the EOS named
\emph{``s95n-v1''} parametrization of Ref.\ \cite{Huovinen:2009yb}, which
is also obtained from lattice QCD \cite{Bazavov:2009zn}. We will use
lattice-QCD data for the magnetic susceptibility $\chi_{m}$ as a function
of $T$ and investigate the effect of the magnetization across the
deconfinement phase transition.

The behaviour of the magnetic susceptibility as a function of the
temperature has been studied in lattice-QCD calculations following
several different approaches [see, \eg
  Refs.\ \cite{Kamikado:2014bua,Bali:2014kia,Bonati:2013vba,Levkova:2013qda}].
Since we concentrate on the deconfinement phase transition and we expect
that $\chi_{m}(T)$ becomes negative in the hadronic phase, we choose the
data from Ref.\ \cite{Bali:2014kia}. Noting that a power-law behaviour
$(T/T_{c}-1)^{n}$ with $n= {\rm const.}$ is typical for quantities near a phase
transition, we capture the analytical behaviour of the data by expressing
$\chi_{m}(T)$ as a polynomial in powers of $T/T_{c}-1$, \ie
\begin{eqnarray}
\chi_{m}(T) & = &
\sum_{n=0}c_{n}\left(\frac{T}{T_{c}}-1\right)^{n}\,,\label{eq:chi_T_fit}
\end{eqnarray}
where $T_{c}$ is the transition temperature and the first six
coefficients appearing in Eq.\ (\ref{eq:chi_T_fit}) are given by
$c_{0}=0.0082$, $c_{1}=0.0374$, $c_{2}=0.0039$, $c_{3}=-0.0427$,
$c_{4}=0.0430$, and $c_{5}=-0.0138$.

Similarly, we introduce the parameter $k_{s}^{2}$, defined as
\begin{equation}
k_{s}^{2}(T) \equiv \frac{p}{e}\,,\label{eq:k_s}
\end{equation}
which is a function of $T$. Note that $k_{s}^{2}$ is reminiscent of, but
distinct from, the square of the sound speed $c^2_s$. The latter is in
fact defined as in Eq. \eqref{eq:cs2} and will in general be different
from the simple ratio of the pressure and energy density [the only
  exception being the EOS of an ultrarelativistic fluid \eqref{eq:EOS_1}
  or the EOS normally used in cosmology $p=we$
  \cite{Rezzolla2013}]. Hence, the parameter $k_{s}^{2}$ does not have a
precise physical significance, but should be seen mostly as a
mathematically convenient definition that allows us to express the
pressure in terms of the energy density and hence obtain an analytic
solution. Having said that, it is interesting also to remark that
$k_{s}^{2}$ and $c^2_s$ are rather similar, as shown in the right panel
of Fig. \ref{fig:Parametrization}, where we compare our fitting functions
for $\chi_{m}$ and $k_{s}^{2}$ with the original data from
Refs.\ \cite{Bali:2014kia} and \cite{Huovinen:2009yb}.

From the \emph{``s95n-v1''} parametrization of
Ref.\ \cite{Huovinen:2009yb}, $p(T)$ and $e(T)$ can be obtained by
integrating over the temperature the trace anomaly $\theta(T) \equiv
e(T)-3p(T)$
\begin{equation}
\label{eq:ep_lattice}
p(T)=T^{4}\int_{T_{1}}^{T}\frac{\theta(T^{\prime})}{T^{\prime5}}dT^{\prime}\,,
\end{equation}
where we choose $T_{1}=1\textrm{MeV}$ and the trace anomaly is given in
Ref.\ \cite{Huovinen:2009yb}
\begin{equation}
\theta(T)=T^{4}
\begin{cases}
  {d_{2}}/{T^{2}}+{d_{4}}/{T^{4}}+{c_{1}}/{T^{n_{1}}} +
  {c_{2}}/{T^{n_{2}}}, & T\geq
  T_{0},\\ a_{1}T+a_{2}T^{3}+a_{3}T^{4}+a_{4}T^{10}, & T< T_{0}\,,
\end{cases}
\label{eq:theta01}
\end{equation}
with $d_{2}=0.2654\,\textrm{GeV}^{2}$,
$d_{4}=6.563\times10^{-3}\,\textrm{GeV}^{4}$,
$c_{1}=-4.370\times10^{-5}\,\textrm{GeV}^{n_{1}}$,
$c_{2}=5.774\times10^{-6}\,\textrm{GeV}^{n_{2}}$, $n_{1}=8$, $n_{2}=9$,
$T_{0}=171.8\,\textrm{MeV}$, $a_{1}=4.654\,\textrm{GeV}^{-1}$,
$a_{2}=-879\,\textrm{GeV}^{-3}$, $a_{3}=8081\,\textrm{GeV}^{-4},$
$a_{4}=-7039000\,\textrm{GeV}^{-10}$. The \emph{ansatz} in
Eq.\ (\ref{eq:theta01}) parametrizes lattice-QCD data when $T\geq T_{0}$,
while it it parametrizes data for a hadron resonance gas when
$T<T_{0}$. Inserting Eqs.\ (\ref{eq:ep_lattice}), (\ref{eq:theta01}) into
Eq.\ (\ref{eq:k_s}), we can get an estimate of $k_{s}^{2}(T)$.

As shown in Fig. \ref{fig:Parametrization}, we also 
obtain the (squared) speed of sound $c_{s}^{2}$ as a function
of $T$ from the \emph{\textquotedblleft s95n-v1''} parametrization of
Ref.\ \cite{Huovinen:2009yb},
\begin{equation}
c_{s}^{2} = \frac{s}{T}\frac{dT}{ds}, 
\label{eq:cs_T_1}
\end{equation}
where entropy density $s\equiv (4p+\theta)/T$ is computed from 
the trace anomaly and Eq.\ (\ref{eq:ep_lattice}).  
%
%
%

Since it is not trivial to obtain the energy density $e$ and the entropy
density $s$ as functions of $T$, we will use the thermodynamical relation
(\ref{eq:Gibbs_03})
\begin{equation}
e=\frac{Ts}{1+k_{s}^{2}}=\frac{s_{0}\tau_{0}}{\tau}\frac{T}{1+k_{s}^{2}}\,,
\end{equation}
where in the second equality we used the conservation of entropy.
Inserting this into the differential equation (\ref{eq:sol_eq_02}),
yields
\begin{equation}
\partial_{\tau}\left[\frac{T}{(1 + k_{s}^{2})\tau}\right] +
\frac{T}{\tau^{2}}-\frac{\tau_{0}}{\tau^{3}}\frac{T_{0}}{1 +
  k_{s}^{2}(T_{0})}\sigma_{0}\chi_{m}=0\,.\label{eq:conser_eq_T_01}
\end{equation}
After solving Eq.\ (\ref{eq:conser_eq_T_01}), we can simply obtain the
energy density via
\begin{equation}
\tilde{e}\equiv\frac{e(\tau)}{e_{0}} =
\frac{\tau_{0}}{\tau}\frac{T(\tau)}{T_{0}}\frac{1 + k_{s}^{2}(T_{0})}{1 +
  k_{s}^{2}(T(\tau))}\,.
\end{equation}

Choosing the initial temperature $T_{0}$, the initial time $\tau_{0}$,
and the critical temperature $T_{c}$ as
\begin{equation}
T_{0}=200\,{\rm MeV}\,,\quad T_{c}=180\,{\rm
  MeV}\,,\quad\tau_{0}=0.5\ {\rm fm/c}\,,
\end{equation}
we solve Eq.\ (\ref{eq:conser_eq_T_01}) numerically and show the results
in Fig.\ \ref{fig:Evolution_02} for different $\sigma_{0}=0,1.0,2.5,25$,
respectively.  Note also that the decay of temperature and energy density
is slowed down due to the magnetization effect. This is consistent with
Eq.\ (\ref{eq:conser_eq_T_01}), which implies that the magnetic field
becomes a source to reheat the system.

As a rough estimate, we can evaluate $\sigma_{0}$ in a typical Bjorken
flow for the QGP. For 200 AGeV Au-Au collisions, the magnetic field can
reach values of $B\sim10\,m_{\pi}^{2}\sim0.44$ GeV, while the energy
density of the magnetic field will be $\frac{1}{2}B^{2}\sim5$ GeV
fm$^{-3}$. The initial energy density of the fluid $e_{0}$ will be
$\sim10$ GeV fm$^{-3}$. Thus we find
$\sigma_{0}=B_{0}^{2}/e_{0}\simeq0.5$, which is close the value for the
blue solid lines in Fig.\ \ref{fig:Evolution_02}.  Overall, this shows
that for a Bjorken flow, the magnetization effect of the QGP can be
ignored. On the other hand, for an ultra-large magnetic field, \eg
$\sigma_{0}=25$ which corresponds to $B\gtrsim50\,m_{\pi}^{2}$, the
temperature will increase, and then decrease, \ie the system is first
reheated by the magnetic field. In that case, which is similar to the one
shown in Fig.\ \ref{fig:03}, the QGP can survive longer than otherwise
expected.

\section{Conclusion}
\label{sec:Conclusion}

We have studied the evolution of the energy density of the QGP produced, for
instance, by the collision of two heavy ions, when this is described as a
one-dimensional, longitudinally boost-invariant flow with a transverse
magnetic field, \ie a transverse Bjorken flow within the ideal-MHD
limit. This represents a rather straightforward extension of our previous
work \cite{Roy:2015kma} to the case in which the flow has a nonzero
magnetization as described via a magnetic susceptibility $\chi_{m}$,
which we have taken to be either constant or to depend on temperature.


Under these conditions for the Bjorken flow, we were able to obtain
analytic solutions relative to two different EOSs, \ie the EOS
(\ref{eq:EOS_1}) for an ultrarelativistic fluid and the EOS
(\ref{eq:EOS_2}) for a magnetized conformal fluid. Interestingly, we find
that all results for a magnetized conformal fluid can be obtained from
the solutions for an ultrarelativistic fluid after a simple
scaling of the susceptibility, \ie by replacing
$\chi_{m}\rightarrow\chi_{m}/3$. We also find that for a paramagnetic
fluid, \ie with $\chi_{m}>0$, the fluid gains energy from the decay of
the magnetic field, thus with an energy density decaying more slowly than
in the case without magnetic fields. On the other hand, for a diamagnetic
fluid, \ie with $\chi_{m}<0$, the fluid loses energy to the magnetic
field and the energy density will decay more rapidly than in the absence
of a magnetic field.

We have also considered the case where the magnetic field is external and
very large, with an evolution that follows a power-law behaviour in
proper time with exponent $a$. The solutions in this case can be
distinguished in terms of two scenarios. The magnetic field decays more
slowly than in the ideal-MHD case for $a>1-\chi_{m}$, while the decay is
more rapid for $a<1-\chi_{m}$.

If the magnetic field or, strictly speaking, $\sigma_{0} =
B_{0}^{2}/e_{0}$, is large enough, the fluid will absorb energy in excess
of the decay caused by the expansion. In that case, one will observe a
peak in the energy density at the early stage. This is a resistive
``reheating'' of the fluid. The amount of this increase depends on the
magnetic-field strength and hence will increase with $\sigma_{0}$ and
$a$. However, at late times, its energy density will decrease with an
asymptotic rate that is the same as in the Bjorken flow, \ie
$\propto\tau^{-4/3}$.

Finally, we have also considered a temperature-dependent magnetic
susceptibility and a realistic equation of state given by lattice-QCD data. We find
that the magnetization effect of the QGP will slow down the decay of
temperature and energy density. However, for realistic values of the
magnetic susceptibility and initial magnetic field, this effect can be
ignored, at least in a one-dimensional Bjorken flow. For an ultra-large
magnetic field, on the other hand, the system might be reheated and the
QGP may survive longer than expected.

\begin{acknowledgments}
S.P.\ and V.R.\ are supported by the Alexander von Humboldt Foundation,
Germany. Partial support comes also from ``NewCompStar'', COST Action
MP1304, by the LOEWE-Program in HIC for FAIR, and by the NSFC under grant
No. 11205150.

Note added -- When this work was being completed, 
we learned that another group, Ref. \cite{Pang:2016yuh}, has investigated 
within a 3+1-dimensional ideal-hydrodynamics approach
 the magnetization effects of an external magnetic field on
 the anisotropic expansion of the QGP.
\end{acknowledgments}

\bibliographystyle{h-physrev}
\bibliography{main}

\end{document}